# Active Perovskite Hyperbolic Metasurface

Zhitong Li[1], Joseph S. T. Smalley[2], Ross Haroldson[3], Dayang Lin[1], Roberta Hawkins[4], Abouzar Gharajeh[1], Jiyoung Moon[4], Junpeng Hou[3], Chuanwei Zhang[3], Walter Hu[1], Anvar Zakhidov[3,5], Qing Gu[1]*

[1]Department of Electrical and Computer Engineering, The University of Texas at Dallas, Richardson, TX, 75080
[2]DigiLens, Inc. 1288 Hammerwood Ave, Sunnyvale, CA 94089
[3]Department of Physics, and [4]Department of Materials Science and Engineering, The University of Texas at Dallas, Richardson, TX, 75080
[5]Department of Nanophotonics and Metamaterials, ITMO University, St. Petersburg, Moscow, Russia
*Corresponding author: qing gu: qing.gu@utdallas.edu

**A special class of anisotropic media – hyperbolic metamaterials and metasurfaces (HMMs) – has attracted much attention in recent years due to its unique abilities to manipulate and engineer electromagnetic waves on the subwavelength scale. Because all HMM designs require metal-dielectric composites, the unavoidable metal loss at optical frequencies inspired the development of active HMMs, where gain material is incorporated to compensate the metal loss. Here, we experimentally demonstrate an active type II HMM that operates at vacuum wavelengths near 750 nm on a silicon platform. Different from previous active HMMs operating below 1 $\mu m$, the dielectric constituent in our HMM is solely composed of gain medium, by utilizing solution-processed and widely tunable metal-halide perovskite gain. Thanks to the facile fabrication, tunability and silicon compatibility of our active HMM, this work paves the way towards HMM's integration into on-chip components, and eventually, into photonic integrated circuits.**

## INTRODUCTION

The electrical response of optical anisotropic medium, in which the electromagnetic wave behavior depends on its polarization state, is mathematically reflected in the tensorial form of its permittivity where not all principle components have the same value [1–3]. Physical phenomena enabled by natural anisotropic materials include dichroism and birefringence, both of which have been thoroughly investigated and used in commercial optical components [4–6]. To further expand anisotropic media's functionalities, metamaterials – artificial materials that have characteristic dimensions much less than their operating wavelengths – provide a fascinating platform to engineer and manipulate permittivities and thus optical properties of materials in dimensions below the diffraction limit.

A unique class of anisotropic media is hyperbolic metamaterials (HMMs), also known as indefinite media, in which the principle components of the effective permittivity tensor have opposite signs. Namely, they exhibit metal and dielectric properties at the same time [7–10]. As a consequence, the isofrequency surface becomes unbounded, i.e. hyperbolic, in the momentum space (k-space), and the optical density of states (DOS) becomes theoretically infinite [11]. Since their inception in 2003 [12], HMMs have been investigated for many applications ranging from super resolution imaging [13–15], to spontaneous emission enhancement [16,17] and topological photonics [18,19]. Although various HMMs have been experimentally demonstrated (See SI part I for a description of HMM classifications), because of the usual metal-dielectric composite nature of optical HMM designs, many of HMM's unique features cannot be accessed because of metal loss at optical frequencies. It is therefore critical to compensate the metal loss in HMMs in order to utilize their full functionalities, and eventually, insert them into optical devices. One approach to compensate loss is to incorporate gain materials, leading to the development of the second generation HMM, also termed active HMM [7,20]. In the visible wavelengths, organic dyes [17,21] and 2D materials [22] have been used as gain due to their low cost and facile fabrication [7]. However, because organic dyes can only be dispersed into dielectrics and 2D materials can only be placed onto HMM surface, the dielectric component is only partially composed of gain in these HMMs, and the resulting loss compensation is not optimal. Recently, full replacement of the passive dielectric by gain was achieved in the telecommunication wavelength in a Ag/InGaAsP multi-layer (i.e. type II HMM) by Smalley et al. [23], in which the active dielectric constituent is an InGaAsP nanostructure defined by e-beam lithography followed by dry etching. Despite the success, this approach has not been adapted at shorter wavelengths, possibly due to the difficulty in constructing deep-subwavelength nanostructures of high aspect ratio in epitaxially grown inorganic gain media. In this work, we demonstrate an active type II HMM in the wavelength range of 740 – 780 nm, utilizing recently emerged metal-halide perovskite gain as the active dielectric component. Not only does such a choice of gain medium allows the construction of nanostructures of narrow width and high aspect ratio [24], many of perovskite's characteristics, including low-temperature solution-processability [25], easy bandgap tunability [26] and large charge-carrier mobility [27], are all desired features of optical gain. Significantly, as in Ref 23 and 29, the optical axis of our active HMM lies tangential to the plane of the substrate. We therefore regard our HMM as an active hyperbolic metasurface, which enables observation of extreme anisotropic behavior for normally incident optical beams.

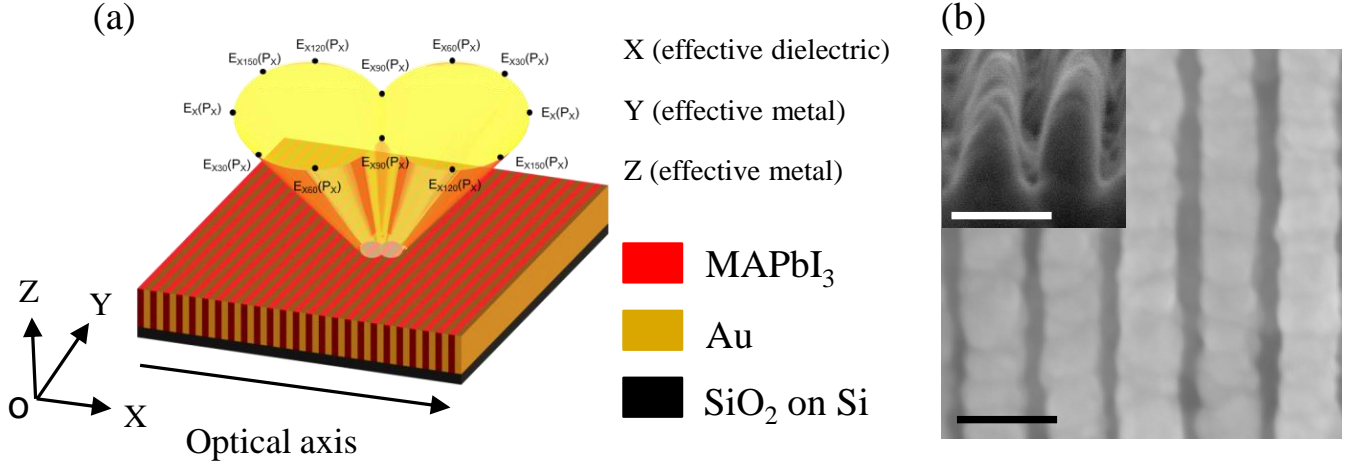

Fig. 1 Type II active MAPbI₃/Au HMM. (a) Schematic of the deep-subwavelength alternating layers of MAPbI₃ (red) and Au (gold) on a Si platform (black). The beam emerging from the HMM surface represents polarization anisotropy, where emission is strong in X direction and weak in Y direction. (b) Top view SEM image of the fabricated active HMM. The scale bar is 100 nm. Inset: Side view SEM image of the active HMM. Note that the top view SEM image is taken at the center of the pattern, while the side view image is taken at the edge.

Our HMM consists of a MAPbI₃ perovskite and Au nanograting superlattice on top of a silicon substrate, in which the dielectric constituent is composed entirely of perovskite. Because stronger hyperbolic behavior is correspondent with stronger anisotropy, we use polarization anisotropy measurements of absorption and emission to quantify the hyperbolic behavior in our active HMMs [23]. We further verify the hyperbolic dispersion in our structure via isofrequency surface analysis and negative refraction simulations. This work marks the first active HMM with operating vacuum wavelength below 1 $\mu m$, where the dielectric constituent is solely composed of gain material. To realize active HMMs in the visible wavelength range, our method can be readily employed by replacing MAPbI₃ with other perovskite compositions of shorter wavelengths. Furthermore, the realization of HMMs on the silicon platform facilitates their integration with optoelectronic components on-chip. Our work opens a path towards HMM's insertion into on-chip devices such as high-speed light sources, electro-optical modulators, and perfect light absorbers, as well as applications such as super resolution imaging and lithography, and topological photonics.

## RESULTS

To design an active HMM with metal-dielectric multilayers, we choose MAPbI₃ among all perovskite compositions due to its longest emission wavelength and in turn, ease of fabrication; we choose Au for its low loss at optical frequencies and minimal chemical reaction with perovskites. The deep-subwavelength MAPbI₃ and Au superlattice is shown in Fig. 1(a). The optical axis is defined along the Bloch vector [28], and the X, Y, and Z directions are defined such that X direction is along the optical axis, while Y and Z directions are orthogonal to the optical axis. To simplify expressions in the proceeding sections, we use notation $E_A$ ($P_B$) to represent the polarization state of emitted (pump) light. For emission polarization $E_A$, the subscript letter A following E denotes the polarization state, and it can be X (polarization state along X direction), Y (polarization state along Y direction), or T (total emission). This subscript is then followed by a number representing the polarization angle with respect to the polarization state direction. Similarly, for pump polarization $P_B$, the subscript letter B following P denotes the polarization state, and it can be either X (polarization state along X direction) or Y (polarization state along Y direction). For example, $E_{X30}$ ($P_Y$) denotes 30 degree-polarized emission under Y-polarized pumping, and $E_T$ ($P_X$) denotes total emission under X-polarized pumping. Lastly, note that $E_{Y30}$ is equivalent to $E_{X120}$ because X and Y directions are orthogonal to each other.

The fabricated active HMM has a period (P) of 80 nm; the Au and MAPbI₃ layers are both 40 nm thick, which is about 5% of the MAPbI₃ photoluminescence (PL) wavelength in the range of 700 nm to 830 nm. Under the effective medium approximation (see SI part II for calculations), the principle components of the effective permittivity tensor satisfy $real(\varepsilon_{xx}) > 0$, $real(\varepsilon_{yy}) < 0$, and $real(\varepsilon_{zz}) < 0$. Namely, the structure behaves as dielectric in X direction and as metal in Y and Z directions. The hyperbolic dispersion of this structure can also be verified using negative refraction simulations (see SI part III for details).

The fabrication of the active HMM consists of three steps: (i). The construction of a deep-subwavelength nanograting Si stamp. (ii). Thermal nanoimprint lithography (NIL) of perovskite using the Si stamp to form a perovskite nanograting. (iii). Thin film evaporation of Au to fill the air trenches of the perovskite nanograting to construct the metal-dielectric multilayer. Fig. 1(b) shows the top- and side- view Scanning Electron Micrograph (SEM) images of the resulting HMM. Note that although the Au deposition is carried out in a planetary sputtering system with high uniformity, air gaps in some patterned areas are unavoidable. Nonetheless, we show that polarization anisotropy and hyperbolic dispersion still exist.

Although etched HMM designs usually assume straight nanograting sidewalls [29], angled sidewalls are inevitable in these structures (Fig. 1(b)). To understand the impact of such fabrication imperfection, we examine the expected anisotropy in realistic structures through transmission and reflection simulations. Fig. 2 shows light transmission through the realistic HMM with angled sidewalls, as functions of the Au fraction at the top and bottom of the superlattice (Au fraction is defined as the Au width divided by the lattice period). In the simulation, the X-polarized incident light propagates from the top to bottom along Z direction, i.e. the effective dielectric direction. Therefore, high transmission and low reflection are expected when the Au and MAPbI₃ ratio falls within the HMM range (Figure S-3), indicating anisotropy. Fig. 2(b) and (c) show transmission as functions of the Au fraction on the top and bottom for an HMM with 80 nm period, while Fig. 2(f) and (g) show the case with 120 nm period. The comparison of the red-colored regions in these two cases illustrates that smaller period corresponds to stronger anisotropy. This is more clearly illustrated in Fig. 2(d) and (h) that define the anisotropic region ($P_X$>0.5 and $P_Y$<0.5), consistent with results of Figure S-5. Therefore, to operate in the anisotropic region, both the top and bottom Au ratios should be close to 0.5. To investigate whether hyperbolic dispersion is achieved in such anisotropic media, isofrequency surface calculation is carried out. The isofrequency surface plots for 80 nm and 120 nm periods are shown in Fig. 2(h). The red line is calculated using the effective medium theory, in which kz is unbounded. The blue line is calculated using layered $MAPbI_3$/Au structure through the transfer matrix theory, in which the largest allowable $kz = \lambda/(P*2\pi)$ is defined by the Bloch theory. In agreement with results of transmission simulations, the largest allowable kz in the structure with 80 nm period (Fig. 2(h) blue) is larger than that with 120 nm period (Fig. 2(h) black), i.e. the superlattice with 80 nm period has a larger region of hyperbolicity. The detailed isofrequency surface analysis with gain can be found in SI Part V (Figure S-10). The reflection and absorption simulation results are provided in Figure S-3.

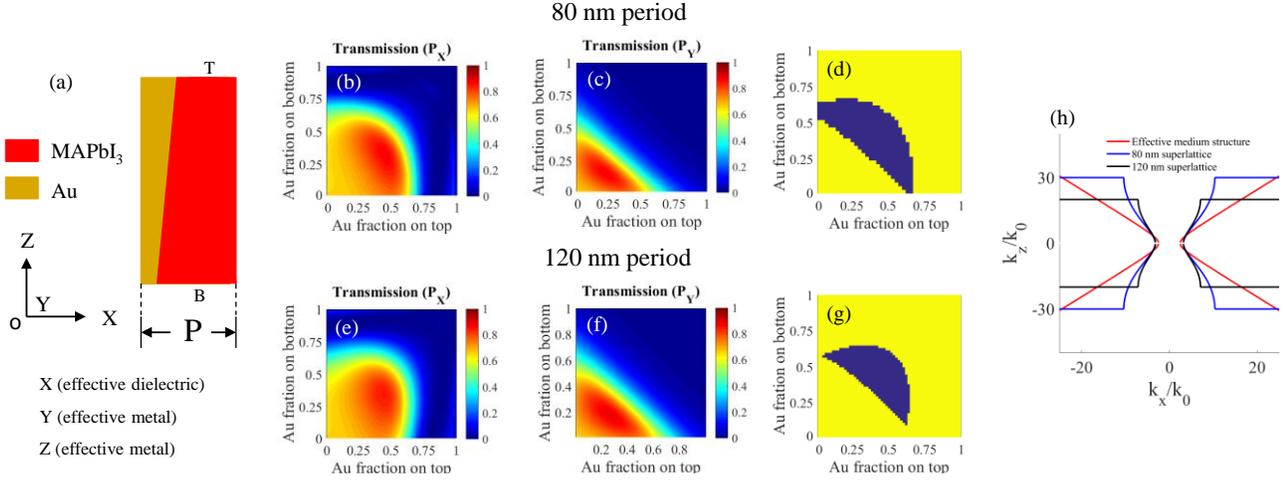

Fig. 2 Simulated polarization anisotropy in the active HMM at 765 nm. (a) Side view schematic of one unit-cell. T and B represent the top and bottom Au widths in one unit-cell, respectively. The top and bottom Au fractions are defined by T/P and B/P, respectively. The period in (b)-(d) is 80 nm. The period in (e)-(g) is 120 nm. (b) Transmission as a function of top and bottom Au fractions with PX. (c) Transmission as a function of top and bottom Au fractions with PY. (d) Identification of the anisotropic region (dark blue). The anisotropic region is defined when transmission is >0.5 with PX and <0.5 with PY. (e) Transmission as a function of top and bottom Au fractions with PX. (f) Transmission as a function of top and bottom Au fractions with PY. (g) Identification of the anisotropic region (dark blue). (h). Real part of the isofrequency surface for 120 nm period HMM. Blue: transfer matrix theory calculation for 80 nm period superlattice. Black: transfer matrix theory calculation for 120 nm period superlattice. Red: effective medium theory calculation.

Given the correlation between anisotropy and hyperbolic dispersion discussed above, i.e. stronger polarization anisotropy is associated with more prominent hyperbolic behavior, we conduct emission and absorption polarization anisotropy measurements to verify the hyperbolic behavior in our structure. Using a micro-PL system (see Methods for details), with the pump light polarized in X direction, we collect PL from the active HMM after its emission passes through a polarizer. As shown in Fig. 3(a)-(b), the highest and lowest emission levels occur at $E_{X0}(P_X)$ and $E_{X90}(P_X)$, respectively. The highest emission level $E_{X0}(P_X)$ has similar intensity to that from a $MAPbI_3$ thin film as well as a $MAPbI_3$ nanograting control samples (see SI part IV), indicating dielectric behavior along X direction. On the other hand, the lowest emission level $E_{X90}(P_X)$ has similar intensity to that from an Au-covered $MAPbI_3$ thin film (i.e. outside of the patterned area as shown in Figure S-7(a)), indicating metal behavior along Y direction. These results are consistent with the effective medium calculation in Figure S-2. Fig. 3(b) depicts the evolution of the integrated emission intensity as a function of the emission polarization angle: it gradually drops from the maximum to the minimum from $E_{X0}(P_X)$ to $E_{X90}(P_X)$, and then increases back to the maximum when the polarization reaches $E_{X180}(P_X)$. We conduct the same measurement on a $MAPbI_3$ thin film control sample

(Figure S-7(b)) and $MAPbI_3$ nanograting (Figure S-8), and only observe slight angle dependent emission, confirming that the large polarization anisotropy is present only in the HMM. To quantify the strength of anisotropy, we calculate the degree of linear polarization (DOLP) for the emitted light and for the absorption of pump light, respectively. Emission DOLP is defined as $((E_X(P_X)-E_Y(P_X))/((E_X(P_X)+E_Y(P_X))$, and a unity DOLP value means linearly polarized emission (i.e. extreme anisotropy), while a zero DOLP value means isotropic emission. As shown in Fig. 3(c), our active HMM's emission is highly anisotropic, with a DOLP value above 0.85 between 740 nm and 800 nm. This value matches well with the simulated emission anisotropy (SI Table 1). On the other hand, absorption DOLP is defined as $(E_T(P_X)-E_T(P_Y))/(E_T(P_X)+E_T(P_Y))$. Fig. 3(d) shows that our HMM has a high DOLP value above 0.8 between 740 nm and 780 nm, exhibiting large absorption anisotropy. Overall, in the fabricated 80 nm period active HMM, strong polarization anisotropy is seen at both the emission (765 nm) and pump wavelength (355 nm).

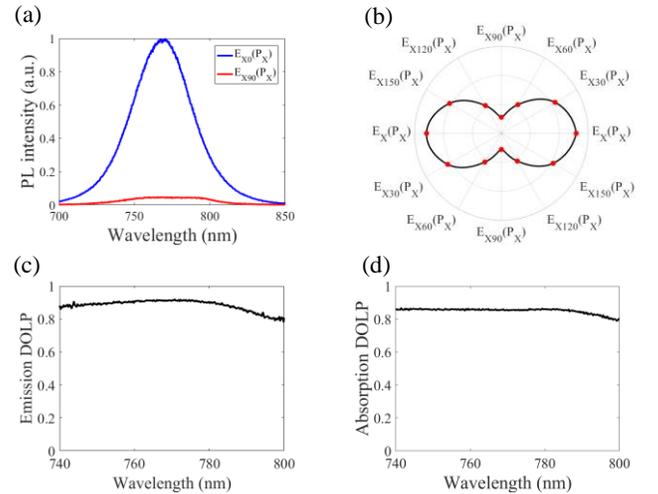

Fig. 3 Measured polarization anisotropy of the active HMM. (a). PL spectra with $E_{X0}(P_X)$ (blue) and $E_{X90}(P_X)$ (red). Note that $E_{Xa}(P_X) = E_{X(180-a)}(P_X)$. (b) Angle dependent total emission integrated from 750 nm to 800 nm. (c) Spectrum of emission DOLP. (d) Spectrum of absorption DOLP.

Fig. 4 shows the experimental comparison of anisotropy in active HMMs of different Au fractions and lattice periods. Comparing HMMs of 0.3 and 0.5 Au fractions with the same 80 nm period, Fig. 4(a-b) show that lower Au fraction corresponds to lower DOLP. Similarly, comparing HMMs of different periods with the same 0.5 Au fraction, Fig. 4(c-d) show that shorter wavelength corresponds to higher DOLP. These results are also verified using the effectiveness examination of effective medium theory (see SI part V).

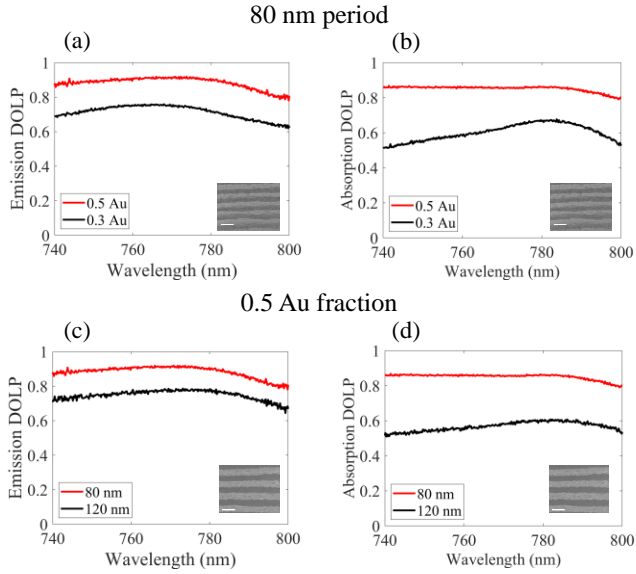

Fig. 4 Measured polarization anisotropy in the active HMM. The inset shows the top view SEM image of the corresponding MAPbI3 nanograting (before metal deposition). (a) Spectrum of emission DOLP with 0.3 Au fraction and 80 nm period (black). Red: Spectrum in Figure 3(c). (b) Spectrum of absorption DOLP with 0.3 Au fraction and 80 nm period (black). Red: Spectrum in Figure 3(d). (c) Spectrum of emission DOLP with 0.5 Au fraction and 120 nm period (black). Red: Spectrum in Figure 3(c). (d) Spectrum of absorption DOLP with 0.5 Au fraction and 120 nm period (black). Red: Spectrum in Figure 3(d).

## CONCLUSION

In summary, we theoretically and experimentally investigated hyperbolic dispersion in an active type II HMM composed of MAPbI3 and Au superlattice on silicon, marking the first active HMM below 1 $\mu m$. Measurements of emission polarization anisotropy, coupled with a systematic theoretical analysis, show the importance of realizing metamaterials describable by effective medium theory to achieve extreme anisotropy. The highly anisotropic behavior of these luminescent HMMs may lead to new strategies towards designing light sources with tailorable polarization, and their realization on the silicon platform opens an avenue towards the integration of HMM into on-chip components such as high speed light sources, electro optical modulators and perfect light absorbers. Thanks to the loss compensation provided by the full replacement of the dielectric constituent with gain, we are a step closer towards HMM's insertion into applications such as super resolution imaging and lithography, and topological photonics.

## METHODS

**Deep-subwavelength Si Stamp Fabrication**: The Si stamp with nanogratings of 80 – 120 nm period is fabricated using standard electron beam lithography and plasma dry etching. A Si substrate is first ultra-sonication cleaned with acetone and IPA for 2 mins each, and its surface is then treated by surpass 3000 (DisChem, Inc.) to increase the adhesion to hydrogen silsesquioxane (HSQ) e-beam resist (Dow Corning@ XR-1541). A 35 nm thick HSQ is then spun-coated on the substrate in two steps (step 1: 500 rpm with 100 rpm/s acceleration for 5 s, step 2: 3000 rpm with 3000 rpm/s acceleration for 60 s), followed by 5 mins of baking on the hotplate at 90 °C to harden the resist. To resolve the deeply subwavelength grating, 1 line with 3.5 nC/cm line dose is written in each period using a RAITH-150 e-beam lithography system, at an acceleration voltage of 30 KV and an aperture size of 30 $\mu m$. After writing the patterns, the sample is developed in 25% TMAH at 38 °C for 1 min. Next, $Cl_2$ plasma dry etching is used to define the Si nanograting with an ICP etcher (Plasma Therm, Inc), with a pressure of 5 mTorr and an ICP power of 500 W. Finally, the HSQ is stripped off in 7:1 buffered oxide etch (BOE) for 90 s.

**Perovskite Thin Film Preparation**: The MAPbI3 solution is made by dissolving a 1:1 molar ratio of $CH_3NH_3I$ and $PbI_2$ in a 7:3 volume ratio of γ-butyrolactone (GBL):dimethylformamide (DMF) solvent mixture in a glovebox. The resulting concentration of the solution is 1.5 M. Next, the solution is heated on a hotplate for 24 hours at 60 °C to fully dissolve. Meanwhile, 1 $\mu m$ $SiO_2$ is thermally grown on a Si substrate, followed by ultra-sonication cleaning with acetone and IPA for 1 min each. Before spin-coating, the $SiO_2$/Si substrate is treated with UV-ozone for 15 mins to increase adhesion between the substrate and MAPbI3. The spin-coating process consists of two steps (step 1. 1000 rpm with 1000 rpm/s acceleration for 23 s, step 2. 4000 rpm with maximum acceleration for 30 s). 12 s into the second spin-coating cycle, 300 $\mu L$ of andydrous toluene is dropped onto the film to increase uniformity. Finally, the sample is annealed at 100 °C on a hotplate for 10 mins.

**Thermal Nanoimprint Lithography of MAPbI3:** The deep-subwavelength Si stamp is first treated by a standard FDTS coating procedure to increase the hydrophobicity on the surface, for subsequent thermal nanoimprint lithography (NIL). The NIL procedure is performed in an Obducat nanoimprinter. We first placed the Si stamp on the spin-coated MAPbI3 perovskite film. To imprint, an aluminum foil is then placed onto the stamp with an O-ring to hold up vacuum. After ensuring that the O-ring is holding the stamp in place, the O-ring is removed and a poly-foil is placed on top of the aluminum foil to increase conformity during NIL. NIL consists of three steps. Step 1: the temperature is gradually increased from 35 °C to 100 °C and the pressure from 0 to 70 bar. The entire process takes 6 mins. Step 2: the temperature and pressure are kept at 100 °C for 20 mins. Step 3: the temperature is gradually dropped from 100 °C to 35 °C in 14 mins. The pressure is kept at 70 bar for the entire 14 mins and is then suddenly released to 0 bar. Finally, the Si stamp is carefully demolded with a blade along the direction of the nanograting.

**Thin Film Evaporation of Au**: After NIL, a 50 nm thick layer of Au is uniformly deposited into the perovskite nanograting in a planetary deposition system (CHA e-gun evaporator) at a chamber pressure of 6×$10^{-7}$ mTorr and with a deposition rate of 0.2 Å/s.

**Theoretical Calculation and COMSOL Simulation:** Theoretical calculations are performed using the effective medium theory (details are provided in SI part II). Transmission and reflection simulations are carried out using an eigenvalue solver in a commercial finite element method (FEM) software COMSOL Multiphysics. Negative refraction simulations are conducted in a frequency domain solver in COMSOL.

**Polarization Anisotropy Characterization:** The polarization anisotropy measurements are conducted at room temperature using a steady-state micro-PL spectroscopy system. The X-polarized 355 nm femtosecond pump beam (Uranus 1030-005-0350-PM, Laser-Femto) is normally incident on the sample under test through a dichroic mirror and an ultraviolet microscope objective with a numerical aperture (NA) of 0.13. A telescope is used to reduce the chromatic aberration by matching the focal planes of the pump and emission wavelengths. To accurately image the pump beam on the HMM pattern, a visible CCD camera (Ophir SP90281) is used in a microscopy configuration. A long-pass dichroic mirror with a cut-off wavelength of 425 nm is introduced to reflect the pump light (355 nm) onto the sample and to pass the emission (765 nm) from the sample. The emitted light is collected by a spectrograph (Princeton Instruments IsoPlane SCT-320) coupled to a cooled Si detector (Princeton Instruments PIXIS:400BRX). In the emission polarization anisotropy measurements, a polarizer is placed in emission path to select specific polarization angles. In the pump polarization anisotropy measurements along Y direction, the X-polarized output beam of the pump laser is rotated to a Y-polarized one by a quarter wave plate followed by a polarizer with its T-axis along Y direction to pass only the Y-polarized beam in the pump path.

**Funding.** This work is supported by the Welch Foundation (grants AT-1992-20190330 and AT-1617), National Science Foundation (No. ECCS-1941629 and CBET-1606141), and The University of Texas at Dallas Office of Research through the SPIRe Grant Program. JH and CZ were also supported by AFOSR (FA9550-16-1-0387), NSF(PHY-1806227), and ARO (W911NF-17-1-0128).

**Disclosures.** The authors declare no conflicts of interest

See Supplement 1 for supporting content.